\documentclass[aps,prd,10pt,tightenlines,notitlepage,nofootinbib,superscriptaddress]{revtex4-1}
\usepackage{amsfonts,amssymb,amsthm,bbm}

\usepackage{amsmath,amssymb,bbm}
\usepackage{amsfonts}
\newcommand{\N}{{\mathbb N}}
\newcommand{\R}{{\mathbb R}}

\newcommand{\cS}{{\mathcal S}}
\newcommand{\SU}{\mathrm{SU}}

\newcommand{\be}{\begin{equation}}
\newcommand{\ee}{\end{equation}}
\newcommand{\beq}{\begin{eqnarray}}
\newcommand{\eeq}{\end{eqnarray}}
\newcommand{\bea}{\begin{eqnarray}}
\newcommand{\eea}{\end{eqnarray}}
\newcommand{\nn}{\nonumber}

\newcommand{\tr}{{\mathrm Tr}}
\newcommand{\f}{\frac}

\def\nn{\nonumber}
\def\pp{\partial}

\def\arr{\rightarrow}

\newcommand{\id}{\mathbb{I}}

\begin{document}

\title{A New Recursion Relation for the 6j-Symbol}

\author{{\bf Valentin Bonzom}}\email{vbonzom@perimeterinstitute.ca}
\affiliation{Perimeter Institute, 31 Caroline St N, Waterloo ON, Canada N2L 2Y5}
\author{{\bf Etera R. Livine}}\email{etera.livine@ens-lyon.fr}
\affiliation{Laboratoire de Physique, ENS Lyon, CNRS-UMR 5672, 46 All\'ee d'Italie, Lyon 69007, France}
\affiliation{Perimeter Institute, 31 Caroline St N, Waterloo ON, Canada N2L 2Y5}

\date{\today}

\begin{abstract}

\noindent The 6j-symbol is a fundamental object from the re-coupling theory of $\SU(2)$ representations. In the limit of large angular momenta, its asymptotics is known to be described by the geometry of a tetrahedron with quantized lengths. This article presents a new recursion formula for the square of the 6j-symbol. In the asymptotic regime, the new recursion is shown to characterize the closure of the relevant tetrahedron. Since the 6j-symbol is the basic building block of the Ponzano-Regge model for pure three-dimensional quantum gravity, we also discuss how to generalize the method to derive more general recursion relations on the full amplitudes.

\end{abstract}

\maketitle

%%%%%%%%%%%%%%%%%%%%%%%%%%%%%%%%%%%%%%%%%%%%%%%%%%%%%%%%%%%%%%%%%%%%%%%%%%%%%%%%%%%%%%%

%%%%%%%%%%%%%%%%%
%\section*{Introduction}
%%%%%%%%%%%%%%%%%

The theory of $\SU(2)$ representations and re-coupling is known for quite a long time. The main motivation was the quantization of angular momenta, with obvious applications in spectroscopy, atomic/molecular physics. Nevertheless, the study of $\SU(2)$ re-coupling coefficients remains an active field of research in modern physics \cite{spinnets-marzuoli}. One reason is the appearance of a new notion to describe some phases of matter: topological order \cite{wen-top-orders}. Its emergence has been described as a condensation of spin networks \cite{levin-wen-condensation} (possibly with a quantum group deformation). It also gives a new way to think of fault-tolerant quantum computations \cite{dennis-top-memory}.

In the same time, a similar formalism has led to interests in re-coupling coefficients to capture aspects of quantum geometry and possibly quantum gravity. The main idea in that approach is that there is no fixed background geometry, but geometry is instead a fully dynamical object. Spin foam models have emerged as direct applications of that idea (see e.g. \cite{SFreview} for a review). A useful model to explore the spin foam program is the Ponzano-Regge model \cite{PR, PR1} which aims at defining the path integral for three-dimensional pure gravity. The latter is a topological field theory: there are no local degrees of freedom due to a large set of specific gauge symmetries. Thus some interests in that model also comes from topological invariants, and especially quantum invariants since a quantum deformation of the Ponzano-Regge model gives the well-defined Turaev-Viro model \cite{TV}.

These models are constructed by triangulating the space-time manifold and defining the corresponding amplitude as a product of local amplitudes attached to each tetrahedron of the triangulation. The basic building block of the Ponzano-Regge model associated to each tetrahedron is the Wigner 6j-symbol of the $\SU(2)$ re-coupling theory.

The 6j-symbol is certainly the most studied Wigner coefficient, and we refer to \cite{qm6j, quantum-tet-marzuoli} for modern reviews describing numerous aspects and to \cite{Varshalovich} for textbook results. An important regime of such a quantum object is the semi-classical limit, when angular momenta become large. The asymptotics of the 6j-symbol has been studied and proved in \cite{SG,roberts, integral,razvan6j}, and next-to-leading orders have been obtained in \cite{NLO_maite, 6jnlo, recursion_maite}. The asymptotics of the Ponzano-Regge model on handlebodies (with more than one tetrahedron) is presented in \cite{dowdall-handlebodies}. Recent works have also focused on more complicated objects, like the 15j-symbol and EPR amplitude in \cite{barrett-asym15j}. An extended Born-Oppenheimer approximation has been developped and applied to the asymptotics of the 9j, 12j and 15j-symbols with some large and small angular momenta \cite{Yu_and_Littlejohn}. One of the author has also studied this regime for arbitrary 3nj-symbols in \cite{3nj-small} with a simpler method.

An efficient way to study such asymptotics as well as algebraic properties of re-coupling coefficients is the use of recursion relations on the angular momenta. It is actually a way to define the 6j-symbol \cite{SG, Varshalovich}, and some insights into asymptotics and recursions on arbitrary 3nj-symbols are given in \cite{3nj-marzuoli}. From the point of view of topological field theory, recursion relations have been understood as a way to encode (some features of) the dynamics, i.e. characterizing the invariance under gauge symmetries \cite{recursion_simone, valentin2}. All these considerations actually also apply to the four-dimensional case \cite{valentin1} and spin foam models for 4d quantum gravity. In this case, we do not yet have the symmetries and the dynamics under full control and we strongly think that the development of recursion relations for the spin foam amplitudes will be a very useful tool to understand further the structure and invariance of 4d spin foams.

There exist several ways to derive recursion relations for the 6j-symbol. The original method is to start from the Biedenharn-Elliott (BE) pentagon identity. The latter encodes at the algebraic level a topological invariance of the Ponzano-Regge model under a special move. It is an equality between the product of two 6j-symbols (the weight for two tetrahedra glued along a triangle) and a sum of products of three 6j-symbols (which correspond to three tetrahedra sharing an edge)\footnote{This states the invariance under the 3-2 Pachner move. Topological invariance also requires invariance under the 4-1 move which is actually induced from the Biedenharn-Elliott identity together with the orthogonality relation of 6j-symbols. But it is only formal due to divergences \cite{PR_john, twisted-homology}.}. Recursion relations are obtained by specializing the BE identity with well-chosen boundary angular momenta \cite{SG}.% One can derive the same recursion relation by simply coming back to the definition of the 6j-symbol from the recoupling of irreducible representations of $\SU(2)$ and compute some insertions of Casimir operators \cite{SG}.

Another path towards recursion relations is to think of Wigner symbols as evaluations of spin network states. Then, one can interpret recursion relations as Wheeler-DeWitt equations in the spin network basis, generated by some Hamiltonian operators which annihilate the evaluations. Such an operator is expected to be the generator of a gauge symmetry \cite{recursion_simone,valentin1,valentin2}. That is a way to see how recursion relations encode and provide the spin network states of Loop quantum gravity with their dynamics. The standard recursion on the 6j-symbol was derived in that way in \cite{valentin2}, where the Hamiltonian operators form a first-class algebra classically, which means that they really generate a gauge symmetry.

The final method which we would like to report on here is to start with the expression of the 6j-symbol as an integral over four copies of $\SU(2)$ (see e.g. \cite{integral}). That group integral approach was already proposed in an earlier study \cite{recursion_simone}. In this previous work, the method could only be applied to a special class of isosceles 6j-symbols (and to the 10j-symbol of the Barrett-Crane model for 4d Euclidean quantum gravity). The idea is that measures in such group integrals have some specific properties which can be used to get recursion formulae. Further, those measures have a geometric nature which eases the interpretation in the asymptotics.

However, the case of generic 6j-symbols has a more complicated measure. Still, it is possible to extract recursion formulae using similar techniques, as we show in the present paper. The organization is the following. In the Section \ref{sec:groupint}, we present the integral which is our starting point. In the Section \ref{sec:newrec}, we prove a new recursion, acting on the square of the 6j-symbol, whose asymptotics and semi-classical interpretation are discussed in the Section \ref{sec:semiclass}. Finally we open in the Section \ref{sec:towards} the interesting possibility of deriving recursion relations in a systematic fashion for arbitrary spin foam amplitudes in the Ponzano-Regge model, as opposed to a single tetrahedron.

%%%%%%%%%%%%%%%%%
\section{6j-Symbol as a Group Integral} \label{sec:groupint}
%%%%%%%%%%%%%%%%%

Let us consider a tetrahedron with its six edges labeled by $i=1,\dotsc,6$ and its four vertices labeled by $a=1,\dotsc,4$. We attach an irreducible representation of $\SU(2)$ to each edge, i.e we associate a spin $j_i\in \N/2$ to the edge $i$. Equivalently, we will refer to these spins as $j_{(ab)}$ where the (symmetric) pair of vertices $a,b$ uniquely defines the edge $i$. At each vertex of the tetrahedron graph, we associate the unique (up to normalization) 3-valent intertwiner between the representations living on the three edges attached to that vertex. The intertwiner is the invariant vector in the tensor product of the three representations meeting at the vertex. Intertwiners are given explicitly by the Clebsch-Gordan coefficients, or more precisely by the Wigner 3jm-symbols. Finally, tensoring these four intertwiners living on the four vertices of the graph and taking the trace on magnetic indices, we obtain the 6j-symbol.

The square of the 6j-symbol can be expressed as an integral over the group $\SU(2)$, using the integral expression of the products of 3jm-symbols \cite{Varshalovich},
\begin{equation}
\int dg\, \prod_{i=1}^3D^{(j_i)}_{a_ib_i}(g)
\,=\,\begin{pmatrix} j_1 &j_2 &j_3\\a_1 &a_2 &a_3\end{pmatrix} \begin{pmatrix} j_1 &j_2 &j_3\\b_1 &b_2 &b_3\end{pmatrix}\;,
\end{equation}
where $D^{(j_i)}(g)$ is the Wigner matrix of $g\in\SU(2)$ in the representation of spin $j_i$ and the quantities into brackets are the 3jm-symbols.
This way, we get one integral over the group for each vertex of the graph. Taking the trace leads to an expression for the square of the 6j-symbol as an integral over $\SU(2)^4$ (see e.g. \cite{integral} for more details),
\be \label{int6jsquare}
\{6j\}^2 \,\equiv\,\begin{Bmatrix} j_{12} &j_{13} &j_{14}\\ j_{34} &j_{24} &j_{23}\end{Bmatrix}^2
\,=\,
\int_{\SU(2)^4} [dg_{a}]^{4}\,
\prod_{(ab)} \chi_{j_{(ab)}}(g_ag_b^{-1})\;,
\ee
where the integral is taken with the Haar measure $[dg]$ over $\SU(2)$ and the character $\chi_j(g)=\tr_j D^{(j)}(g)$ is the trace of the Wigner matrix $D^{(j)}(g)$. This is the starting point of our analysis.

Since characters are invariant under conjugation by $\SU(2)$ group elements, we can write the above expression as an integral over the class angles $\theta_{ab}$ of the group elements $g_ag_b^{-1}$. The change of measure was computed in \cite{integral},
\be \label{angleint}
\{6j\}^2
\,=\,
\f2{\pi^4}\int_{D} [d\theta_{ab}]^{6}\,
\f1{\sqrt{\det_{4\times 4} (\cos\theta_{ab})}}\,
\prod_{(ab)} \sin (2j_{ab}+1)\theta_{ab}\;,
\ee
where we take the convention that $\theta_{aa}=0$ and thus $\cos\theta_{aa}=1$. The measure is given in term of the determinant of a $4\times 4$ Gram matrix $(\cos\theta_{ab})$. It has a simple geometrical interpretation: ${\sqrt{\det (\cos\theta_{ab})}}$ gives the volume of the (dual) tetrahedron on the sphere $\cS^3$, whose spherical edge lengths are the angles $\theta_{ab}$. The above formula can be used to extract the asymptotic behavior of the 6j-symbol at large spin through a saddle point analysis \cite{integral}.

The domain of integration $D$ is the set of all possible spherical tetrahedra, i.e. the subset of $[0,\pi]^6$ satisfying $\theta_{ac}\leq \theta_{ab}+\theta_{bc}$ and $\theta_{ab}+\theta_{bc}+\theta_{ca}\leq 2\pi$.

Since it will be our departure point, let us sketch the derivation of \eqref{angleint}. Among the four group variables in \eqref{int6jsquare}, we can absorb, say, $g_4$ in a re-definition of the others, $g_a\mapsto g_a g_4^{-1}$ for $a=1,2,3$. Clearly, that does not change the arguments of the characters. Moreover, due to the right invariance of the Haar measure, $dg_1, dg_2, dg_3$ are unchanged, and the integral over $g_4$ becomes trivial. Then we write each group element $g_a = \exp(i \theta_{a4}\, \hat{n}_{a}\cdot\vec{\sigma})$, where $\theta_{a4}$ is the class angle, and $\hat{n}_a\in\cS^2$ the axis of the rotation, and the measure reads
\be
dg_a=\frac{2}{\pi}\ \sin^2\theta_{a4} d\theta_{a4}\,d^2\hat{n}_a\;.
\ee
The class angles are kept as integration variables since they are the arguments of the characters with spins $j_{a4}$. As for the six variables $(\hat{n}_a)_{a=1,2,3}$, we split them into two sets. One consists in the dot products $u_{ab}\equiv \hat{n}_a\cdot\hat{n}_b$, and the other set form a global rotation (three components) which leaves the variables $u_{ab}$ invariant and trivially factorizes. Finally, we notice that the arguments of the characters with spins $(j_{ab})$ with $a,b\neq 4$ are class angles, which we denote $\theta_{ab}$, which are related to the variables $(u_{ab})$ by
\be
\cos \theta_{ab} = \cos\theta_{a4}\,\cos\theta_{b4} + \sin\theta_{a4}\,\sin\theta_{b4}\,u_{ab}\;.
\ee
The final step is to change the variables from $u_{ab}$ to $\theta_{ab}$. The Jacobian is straightforward to evaluate and leads to \eqref{angleint}.

%%%%%%%%%%%%%%%%%
\section{New Recursion} \label{sec:newrec}
%%%%%%%%%%%%%%%%%

Let us introduce the following notations,
\be
P[X_{ab}]\equiv \det(X_{ab}),\quad \text{and}\qquad T^\pm_{ab} f(j_{ab}) = f(j_{ab}\pm\tfrac12)\;.
\ee
$P$ is the determinant of a collection of numbers $(X_{ab})$, and is polynomial of degree four in the six variables $X_{ab}=X_{ba}$, $X_{aa}=1$. The operators $T^\pm_{ab}$ shift the spin $j_{ab}$ for a given pair $(ab)$. The new recursion we are going to present is
\be \label{rec6jsquare}
\left[4\,
P[\tfrac12(T^+_{ab}+T^-_{ab})]\,
\left(
(2j_{\alpha\beta}+2)\,T^+_{\alpha\beta}\,-\,2j_{\alpha\beta}\,T^-_{\alpha\beta}
\right)
+ \pp_{X_{\alpha\beta}}P[\tfrac12(T^+_{ab}+T^-_{ab})]\,(T^+_{\alpha\beta}-T^-_{\alpha\beta})^2
\right]\
\{6j\}^2=0 \;,
\ee
for each symmetric pair $(\alpha\beta)$ of vertices, with $\alpha,\beta=1,2,3,4$ and $\alpha\neq\beta$.

Our method is fairly simple and works exactly like Schwinger-Dyson equations in field theory. We multiply the integrand of \eqref{angleint} with a well chosen insertion and take a total derivative and integrate over all angles.

Singling out one edge $(\alpha\beta)$, we consider the following integral
\be
\int [d\theta_{ab}]^{6}\
\pp_{\theta_{\alpha\beta}}
\Bigl[
\,\sin\theta_{\alpha\beta}\,\sqrt{\det_{4\times 4} (\cos\theta_{ab})}\,
\prod_{(ab)} \sin (2j_{ab}+1)\theta_{ab}
\Bigr]
\,=\,
0\;,
\ee
which is indeed zero since the factor $\det(\cos\theta_{ab})$ vanishes on the boundary of the domain $D$ \cite{integral}.

To expand this integral, we use the notation $P[X_{ab}]= \det_{4\times 4} X_{ab}$. The measure term is given by its  evaluation on $X_{ab}=\cos\theta_{ab}$. We compute explicitly the derivative in the previous expression and distinguish the two terms where we differentiate the Gram matrix determinant or the other terms,
\beq
0&=&
\int [d\theta_{ab}]^{6}\,\,
%\f{1}{\sqrt{\det (\cos\theta_{ab})}}\,
\Bigg{[}\f{\det (\cos\theta_{ab})}{\sqrt{\det (\cos\theta_{ab})}}\,
\Bigl(
\cos\theta_{\alpha\beta} \,\sin(2j_{\alpha\beta}+1)\theta_{\alpha\beta}
+(2j_{\alpha\beta}+1) \,\sin\theta_{\alpha\beta} \,\cos(2j_{\alpha\beta}+1)\theta_{\alpha\beta}
\Bigr) \nn\\
&&-\,\f{\pp_{X_{\alpha\beta}}P[X_{ab}]|_{X=\cos\theta}}{2\sqrt{\det (\cos\theta_{ab})}}
\,\sin^2\theta_{\alpha\beta} \,\sin(2j_{\alpha\beta}+1)\theta_{\alpha\beta}\,\Bigg{]}
\prod_{(ab)\ne(\alpha\beta)} \sin (2j_{ab}+1)\theta_{ab}\;,
\eeq
which can be re-written as
\beq
0&=&
\int [d\theta_{ab}]^{6}\,\,
%\f{1}{\sqrt{\det (\cos\theta_{ab})}}\,
\Bigg{[}\f{\det (\cos\theta_{ab})}{\sqrt{\det (\cos\theta_{ab})}}\,
\Bigl(
(2j_{\alpha\beta}+2)\,\sin(2j_{\alpha\beta}+2)\theta_{\alpha\beta}
-2j_{\alpha\beta}\,\sin 2j_{\alpha\beta}\theta_{\alpha\beta}
\Bigr) \nn\\
&&-\,\f{\pp_{X_{\alpha\beta}}P[X_{ab}]|_{X=\cos\theta}}{\sqrt{\det (\cos\theta_{ab})}}\,
(1-\cos^2\theta_{\alpha\beta}) \,\sin(2j_{\alpha\beta}+1)\theta_{\alpha\beta}\,\Bigg{]}
\prod_{(ab)\ne(\alpha\beta)} \sin (2j_{ab}+1)\theta_{ab}\;.
\eeq
Now all the $\cos\theta$-factors can be re-absorbed as shifts in the spin labels $j_{ab}$. Indeed, the multiplication by a factor $\cos\theta_{ab}$ corresponds to the operator $\f12(T^+_{ab}+T^-_{ab})$ due to the trigonometric identity
\be \label{trigo}
\cos\theta_{ab}\ \sin (2j_{ab}+1)\theta_{ab}
\,=\,
\frac{\sin (2j_{ab}+2)\theta_{ab}
\,+\,
\sin (2j_{ab})\theta_{ab}}2 \,=\, \frac{T^+_{ab}+T^-_{ab}}{2}\ \sin (2j_{ab}+1)\theta_{ab} \,.
\ee
That leads to the following recursion relation on the squared 6j-symbol
\be \nonumber
\left[4\,
P[\tfrac12(T^+_{ab}+T^-_{ab})]\,
\left(
(2j_{\alpha\beta}+2)\,T^+_{\alpha\beta}\,-\,2j_{\alpha\beta}\,T^-_{\alpha\beta}
\right)
+ \pp_{X_{\alpha\beta}}P[\tfrac12(T^+_{ab}+T^-_{ab})]\,(T^+_{\alpha\beta}-T^-_{\alpha\beta})^2
\right]\
\{6j\}^2=0 \;,
\ee
where we used that $T^+_{\alpha\beta}T^-_{\alpha\beta}=T^-_{\alpha\beta}T^+_{\alpha\beta}=\id$. This is \eqref{rec6jsquare}. The non-triviality of the relation is ensured by the presence of the factors $(2j+2)$ and $2j$ in the first line, which implies that the coefficients of this equation can not all vanish.

\medskip

To better understand the explicit structure of this relation, we need to expand the Gram determinant,
\be
P[X_{ab}]=\sum_{\sigma\in{\cal S}_4}\epsilon(\sigma)\,\prod_a^4X_{a\sigma(a)}\;,
\ee
where we sum over all permutations of 4 elements. We classify them by their number of fixed points. The only permutation that fixes all elements is the identity. Then we get the two-point cycles, which exchange two elements without touching the other two elements. These contribute as $-X_{ab}^2$ to the determinant, where $a$ and $b$ are the two elements which are interchanged. Then permutations with a single fixed point are given by the 3-cycles, e.g. $1\arr 2\arr 4\arr 1$ leaving $3$ invariant. These give terms of order 3. Finally, we have the six 4-cycles, which lead to term of order 4 in the polynomial. More explicitly, we can compute
\beq
P[X] &=& 1- X^2_{12}- X^2_{13}- X^2_{14}- X^2_{23}- X^2_{24}- X^2_{34} \nn\\
&& +2X_{23}X_{24}X_{34}+2X_{13}X_{34}X_{14}+2X_{12}X_{24}X_{14}+2X_{12}X_{23}X_{13}\nn\\
&&+X_{12}^2X_{34}^2+X_{13}^2X_{24}^2+X_{14}^2X_{23}^2\nn\\
&&-2X_{12}X_{23}X_{34}X_{14}-2X_{13}X_{34}X_{24}X_{12}-2X_{14}X_{24}X_{23}X_{13}\;,
\eeq
with $4!=24$ terms. We also compute its derivative with respect to, say, $X_{12}$,
\be
\pp_{X_{12}}P
\,=\,
2\left(X_{12}(X_{34}^2-1)+X_{23}X_{13}+X_{24}X_{14}-X_{23}X_{14}X_{34}-X_{24}X_{13}X_{34}
\right)\;.
\ee
In the end, we obtain a recursion relation which involves mixed shifts on the spins $j_{ab}$ up to five $\pm\f12$-shifts.

%%%%%%%%%%%%%%%%%
\section{Asymptotics and Semi-Classical Geometry} \label{sec:semiclass}
%%%%%%%%%%%%%%%%%

%%%%%%%%%%%%%%%%%
\subsection{The Recursion as a Closure Condition}
%%%%%%%%%%%%%%%%%

The most well-known (and useful) recursion relation on the 6j-symbol is a second order difference equation \cite{Varshalovich},
\be \label{rec6j}
A_{+1}[j_i]\,\begin{Bmatrix} j_1+1 &j_2 &j_3 \\ j_4 &j_5 &j_6 \end{Bmatrix} +
A_{0}[j_i]\,\begin{Bmatrix} j_1 &j_2 &j_3 \\ j_4 &j_5 &j_6 \end{Bmatrix} +
A_{-1}[j_i]\,\begin{Bmatrix} j_1-1 &j_2 &j_3 \\ j_4 &j_5 &j_6 \end{Bmatrix} = 0,
\ee
and there also exists a similar second order recursion with $\pm \f12$-shifts. The explicit coefficients can be found in \cite{Varshalovich,SG} for instance. This equation can be used to derive all 6j-symbols from an initial condition, and is thus used numerically to generate lists of values of the symbol. This means that in principle the recursion relation \eqref{rec6jsquare} could be obtained from this one. This seems however to be quite complicated.

Still it is interesting to compare both:
 \begin{itemize}
 \item the standard relation only has three terms, but complicated coefficients,
 \item while our new recursion relation has many more terms (corresponding to the various combinations of shifts), but all coefficients are trivial (either $\pm1$ or $\pm(2j+2)$, or $\pm 2j$).
 \end{itemize}
Hence it certainly can not be used in efficient numerical simulations, but in contrast its geometrical interpretation at leading order is straightforward, as we will now explain.

In geometric terms, the recursion \eqref{rec6j} has been interpreted in \cite{recursion_simone} as the statement of invariance under an elementary deformation on a tetrahedron whose result is a small displacement of a vertex. This simply comes out of interpreting some specialization of the Biedenharn-Elliott pentagon identity \cite{Varshalovich} in the 3-2 Pachner move. From this point of view, the recursion \eqref{rec6jsquare} should certainly be seen as a non-trivial combination of such elementary moves, successively performed. The question is then what geometric properties are encoded into such combination of moves. While the standard recursion \eqref{rec6j} encodes the full symmetry of the model (topological invariance), the new recursion should only probe some simpler geometric properties of the 6j-symbol. To extract this information, it is far easier and more fruitful to look at its asymptotic form. At the leading order, this is just the closure of the corresponding tetrahedron.

The asymptotics of the 6j-symbol when all spins are homogeneously scaled is\footnote{Note that it can be efficiently derived from \eqref{rec6jsquare}.} (see \cite{roberts,PR,SG,integral})
\be
\{6j\}\sim\frac{\cos\bigl(S_R[j_i]+\f\pi 4\bigr)}{\sqrt{12\pi V[j_i]}}\;.
\ee
Here, $V[j_i]$ is the volume of the tetrahedron  with edge lengths $j_i+\f12$ and $S_R[j_i]=\sum_i (j_i+\f12)\phi_i[j_i]$ is the Regge action for the same tetrahedron. The angles $\phi_i[j_i]$ are the dihedral angles of the tetrahedron computed as functions of the edge lengths. In particular, they satisfy the following property,
\be
\det_{4\times 4}(\cos \phi_{ab}) = 0\;.
\ee
Mathematically, it means exactly that the angles $\phi_{ab}$ are dihedral angles for some tetrahedron. Equivalently, we will refer to that condition as the \emph{closure} of a tetrahedron. This is part of the information on the $\phi_{ab}[j_i]$. The other information that is needed to recover the asymptotics is that they are the dihedral angles for the tetrahedron whose lengths are $(j_{ab}+\frac12)$. Those two conditions are the roots of the first order formulation of Regge calculus, where lengths and angles are independent variables \cite{first-order-regge}. They can be both extracted from the recursion \eqref{rec6j}, \cite{SG}.

Squaring the asymptotic approximation, we easily get
\be
\{6j\}^2\sim\f1{24\pi V[j_i]}\,\left(1-\sin\bigl(2\,S_R[j_i]\bigr)\right),\quad \text{where}\qquad
2S_R[j_i]=\sum_i (2j_i+1)\phi_i[j_i]\;.
\ee
This leading order asymptotics goes as $j^{-3}$ as we scale the spins homogeneously.

Let us insert it in the new recursion relation and check that it solves it at leading order in the large spin limit. Ignoring the volume pre-factor (which only contributes to higher orders), we have two terms in the asymptotics: the constant term $1$ and the oscillatory term $\sin(2S_R[j_i])$.

The constant term trivially solves the recursion relation because of the perfect balance of plus and minus signs in the equation. More precisely, both the shifts due to the Gram determinant (since $\sum_\sigma \epsilon(\sigma)=0$) and due to the $(1-\cos^2\theta)$ factor (and actually also the shifts due to $\pp_X P$) vanish.

The non-trivial check thus comes from the oscillatory term. Let us look at the behavior of $\sin 2S_R$ when spins are shifted. By applying basic trigonometry, we easily see that
\beq
(T^+_{ab}+T^-_{ab})\ \sin (2S_R[j_i])
&=& 2\cos\phi_{ab}\,\sin (2S_R[j_i])\;,\nn\\
(T^+_{ab}-T^-_{ab})\ \sin (2S_R[j_i])
&=& 2\sin\phi_{ab}\,\cos(2S_R[j_i])\;.
\eeq
Keeping only the leading contribution to the recursion relation, i.e. only the terms with an explicit $j_{\alpha\beta}$, we see that it holds if and only if
\be
%\cos\phi_{\alpha\beta}\,\det_{4\times 4}\cos\phi_{ab}
%\,=\,
P[\tfrac12(T^+_{ab}+T^-_{ab})]\,\bigl(T^+_{\alpha\beta}-T^-_{\alpha\beta}\bigr)\,\sin(2S_R[j_i])\propto
\sin\phi_{\alpha\beta}\,\det_{4\times 4}\cos\phi_{ab}\ \cos(2S_R[j_i])
\,=\, 0\;.
\ee
This is obviously true due to the closure of the tetrahedron, which directly implies the vanishing of the Gram determinant of the geometrical dihedral angles, $\det_{4\times 4}\cos\phi_{ab}=0$.

Equivalently, if we try to find solutions to the new recursion \`a la WKB, we find that for any oscillations going like $\exp(i\sum j_{ab}\, \omega_{ab}[j_i])$, the angles $\omega_{ab}$ are constrained to be dihedral angles for some tetrahedron, whatever they are as functions of the spins.

If we want to push the analysis to higher orders, we have to include the next-to-leading order the 6j-symbol (e.g. \cite{recursion_maite,NLO_maite}) and the corrections due to the volume pre-factor. The geometrical interpretation of the higher order corrections to the 6j-symbol itself is not clear at all, so it might be enlightening to push this analysis and see if our new recursion relation can provide a natural geometrical meaning. The hope is that our recursion does not encode the full symmetries of the 6j-symbol but focuses on its closure which maybe partially characterize the next-to-leading orders of the 6j-symbol.

%%%%%%%%%%%%%%%%%
\subsection{Comparison with other Closure Equations}
%%%%%%%%%%%%%%%%%

%%%%%%%%%%%%%%%%%
\subsubsection{Asymptotic Equations} \label{sec:asym}
%%%%%%%%%%%%%%%%%

One of the key idea in loop quantum gravity and spin foam models is to make sense of quantum geometry using $\SU(2)$ re-coupling. From this view point, it is expected that the relevant $\SU(2)$ objects satisfy such closure equations.
So far, it has been mainly investigated in the context of models for four-dimensional quantum geometry. It has been recently shown in \cite{valentin1} that closure relations generically holds asymptotically in the large spin limit when the leading order oscillates with the Regge action of the 4-simplex.

Consider a $d$-simplex (with $d=3,4$) with its $(d-2)$-subsimplices labeled by spins $(j_{ab})$. Interpreting the latter as the volumes of the $(d-2)$-simplices, we assume we are in a regime where a classical geometry of the simplex with those volumes is available. This assumption is obviously satisfied for the 6j-symbol, but is more subtle in four dimensions since ten areas do not always determine a flat 4-simplex. It can be reached nevertheless by using an overcomplete basis of intertwiners which typically provides the normals to the triangles in addition to their areas. The assumption then implies that dihedral angles are well-defined and depend on the spins. They may also depend on the normals but we can neglect their variations to leading order.

If the amplitude goes like $W[j_{ab}]\sim A\,\exp(iS_R[j_{ab}])$, then it satisfies the following equation for each spin $j_{ab}$, \cite{SG, valentin1},
\be \label{regge-criterion}
\left((T^+_{ab})^2+(T^-_{ab})^2\right)\,W[j_{ab}] \,=\, 2\,\cos\phi_{ab}\,W[j_{ab}]\;.
\ee
It means that multiplication by the cosine of a dihedral angle can be compensated by shifts on the corresponding spin of $\pm 1$. Then taking the antisymmetrized sum over all permutations of products of that equation, we form on the right hand side an identically vanishing factor, $\det(\cos \phi_{ab}) =0$, so that
\be \label{asymclosure}
P[(T^+_{ab})^2+(T^-_{ab})^2]\ W[j_{ab}]\,=\,0\;.
\ee
That provides a definition of a closure operator.

%\footnote{It exists obviously for the 6j-symbol, but it is subtler in four dimensions, since ten areas do not always determine a flat 4-simplex. Nevertheless, a classical geometry can be reached by using an overcomplete basis of intertwiners which gives actually more information than just the volumes of the $(d-2)$-simplices.} implies that dihedral angles are functions of the spins, $\phi[j_{ab}]$. If the amplitude goes like $W[j_{ab}]\sim A\,\exp(iS_R[j_{ab}])$.

%%%%%%%%%%%%%%%%%
\subsubsection{Exact Equations}
%%%%%%%%%%%%%%%%%

The above relation only holds asymptotically, for the 6j-symbol, the 15j-symbol and more complicated objects \cite{valentin1}. Hence getting an \emph{exact} closure relation for the 6j-symbol is an interesting achievement.

The technical difference between \eqref{asymclosure} and the asymptotic form of the exact equation \eqref{rec6jsquare} is the following. The operator $(T^+_{ab})^2+(T^-_{ab})^2$ which is the argument of $P$ in \eqref{asymclosure} generates shifts of $\pm1$ while in \eqref{rec6jsquare}, the argument of $P$ is $\frac12 (T^+_{ab}+T^-_{ab})$ which generates shifts of $\pm\frac12$. That difference is compensated since $P$ acts on the 6j-symbol itself in \eqref{asymclosure} but on its square in the exact recursion. Note also that the exact recursion picks up additional terms in the sub-leading orders, due to the derivatives of $P$ in \eqref{rec6jsquare}.

The new recursion \eqref{rec6jsquare} is actually the second known example of an exact relation whose leading asymptotic behavior reduces to the closure of a simplex. The first example we have found is the 10j-symbol. The similarities of the derivation with the present situation strengthens the potential generality of the method. Hence we briefly sketch a few steps of that derivation which was reported with details in \cite{recursion_simone}.

The 10j-symbol is a quantum weight associated to a 4-simplex in a model for four-dimensional geometry known as the Barrett-Crane spin foam model \cite{BC}. It is a function of ten spins $j_{(ab)}$ living on the 4-simplex boundary graph (the complete graph with 5 vertices, where each vertex is linked to the other four vertices by a single edge) and its asymptotics displays oscillations with the Regge action of the 4-simplex \cite{asym10j}. It is simply expressed as an integral over five copies of $\SU(2)$,
\be \label{10jgroup}
\{10j\}
\,=\,
\int_{\SU(2)^5} [dg_{a}]^{4}\,
\prod_{(ab)} \chi_{j_{ab}}(g_ag_b^{-1})\;.
\ee
Once again, we can write this as an integral over the class angles of the group elements $g_ag_b^{-1}$ and the structure of the 10j-symbol is actually simpler than that of the squared 6j-symbol \cite{integral}:
\be \label{10jangles}
\{10j\}
\,=\,
\f4{\pi^6}\int_0^{\pi} [d\theta_{ab}]^{10}\,
\delta(\det_{5\times 5} (\cos\theta_{ab}))\,
\prod_{(ab)} \sin (2j_{ab}+1)\theta_{ab}\;,
\ee
where we simply have to impose that the $5\times 5$ Gram matrix has a vanishing determinant. This constraint states that the ten angles $\theta_{ab}$ are the dihedral angles of a true geometric, flat 4-simplex.

To get a recursion, we insert the determinant of the Gram matrix in the integral, which then vanishes due to the constraint in the measure term,
\be
\int [d\theta_{ab}]^{10}\
\delta(\det_{5\times 5} (\cos\theta_{ab}))\,
\det_{5\times 5} (\cos\theta_{ab})\,
\prod_{(ab)} \sin (2j_{ab}+1)\theta_{ab}
\,=\,
0\;.
\ee
The new factor $\det_{5\times 5} (\cos\theta_{ab})$ is a polynomial in the variables $\cos\theta_{ab}$. Each cosine actually produces a $\pm\f12$-shift in the $j_{ab}$'s spins labeling the 10j-symbol, due to the trigonometric identity \eqref{trigo}.

In \cite{recursion_simone}, it was checked that amplitudes going like $\exp(i\sum_{(ab)} (2j_{ab}+1)\phi_{ab})$, i.e. which oscillates with the Regge action of the 4-simplex whose triangle areas have the values\footnote{The 10j-symbol has only ten quantum numbers which do not always determine a 4-simplex. But it is known that there is a regime where it does \cite{asym10j}.} $(j_{ab})$, do indeed satisfy the above equation. That is because, not very surprisingly, it reduces to $\det(\cos\phi_{ab})$ in the asymptotics.

The surprising fact about that equation on the 10j-symbol is that it is exact although the Regge oscillations are definitely not the dominant contributions to the asymptotics \cite{nloBC}. This is in contrast with the case of the 6j-symbol we have been studying here. The 10j-symbol is thus a good example to illustrate the difference between the reasoning of the Section \ref{sec:asym} in the large spin limit and the exact treatment (when available). In particular, we do not expect the 10j-symbol to satisfy the equation \eqref{regge-criterion} at the leading order in the asymptotics.

%%%%%%%%%%%%%%%%%
\section{Towards Recursion Relations for Spinfoam Amplitudes} \label{sec:towards}
%%%%%%%%%%%%%%%%%

%We are not sure of the relation between our new recursion relation which involves $\pm\f12$-shifts on all spins and the more standard recursion relation which involves $\pm 1$-shifts on a single spin (or $\pm\f12$-shifts on three spins). The standard relation only has three terms, but complicated coefficients. It can used to compute efficiently 6j-symbols. Our new recursion relation has many more terms corresponding to the various combinations of shifts, but all the coefficients are trivial (either $\pm1$ or $\pm(2j+2)$ or $\pm 2j$). It can not be used that  efficient numerical simulations, but its geometrical interpretation at leading order is straightforward as we have explained above.

%\medskip

%Another big difference between the present approach and the standard method is that the usual recursion relation deals directly with the  6j-symbol, while our method applies to the square of the 6j-symbol. While this makes it hard to see how the two are effectively related, it also opens the possibility that the geometrical interpretation of the higher order asymptotics of the squared 6j-symbol be simpler than for the 6j-symbol itself. We will explore this possibility elsewhere.

The previous sections were dedicated to the new recursion from the point of view of the 6j-symbol itself. Here we take a different perspective and try to sketch some implications for the Ponzano-Regge model \cite{PR,PR1}.

An important difference between the standard recursion \eqref{rec6j} and the new one \eqref{rec6jsquare} that we want to emphasize is that the latter acts on the square of the 6j-symbol, as opposed to the 6j-symbol itself. While the 6j-symbol is the building block of the Ponzano-Regge model, which is attached to each tetrahedron of a triangulation, its square is the amplitude for the 3-sphere triangulated by two tetrahedra which are identified along their boundary,
\be
Z_{\rm PR}(\cS^3) = \int \prod_{a=1}^4 dg_a \ \prod_{a<b} \delta(g_a\,g_b^{-1}) = \sum_{\{j_{ab}\in\frac{N}{2}\}_{a<b}} \Bigl[\prod_{a<b} (2j_{ab}+1)\Bigr]\ \begin{Bmatrix} j_{12} &j_{13} &j_{14}\\ j_{34} &j_{24} &j_{23}\end{Bmatrix}^2\;.
\ee
This equation is only formal since it is actually divergent (there is one redundant Dirac delta in the product over all $a<b$). Making sense of that expression normally involves a gauge-fixing process, described in \cite{PR1}, which amounts to setting one spin in the summand to zero.

Our new recursion allows a different analysis (which we unfortunately cannot make complete). Note that the standard recursion on the 6j-symbol \eqref{rec6j} cannot be used in the above formula, since it requires to transform independently each symbol. Hence it is not a consistent symmetry of the amplitude. The new recursion, however, is directly a symmetry of the summand.

The usual recursion \eqref{rec6j} encodes the gauge invariance of the model. From the spacetime point of view, that comes from specializing the Biedenharn-Elliott identity which states the invariance under the 2-3 Pachner move. In the Hamiltonian framework, it is a quantization of a generator of a gauge symmetry which enables to solve for the wave-function on the boundary of a tetrahedron \cite{valentin2}. So it may be tempting to interpret the new recursion as being also related to that gauge symmetry. However, that is unclear to us, and we have looked at the algebra of the recursions acting on different spins without finding pieces of evidence.

Instead, the new recursion can be interpreted as a \emph{global} symmetry of the summand. By global, we mean that it acts on all spins, none of them being left unchanged. Together with the result of the previous section, that gives the following picture. The shifts on the squared 6j-symbol correspond to a deformation of the triangulation of the 3-sphere equipped with a discrete metric (given by the arguments of the symbol). Invariance under that deformation implies that the metric has to close the boundaries of the tetrahedra (which are identified). Thus, it provides with a consistent metric deformation of the triangulation.

%From this point of view, such deformations could actually be interpreted as discrete diffeomorphisms acting on the space-time triangulation and then our new recursion relation would represent the invariance under these discrete diffeomorphisms.

\medskip

We would like to propose to develop this perspective and write recursion relations for more general triangulations. Let us start with a triangulation $\Delta$ of the 3-sphere. We consider a connected graph $\Gamma$ on $\Delta$, and fix all the spins $j_e$ on the edges $e\in\Gamma$ to fixed values. Looking at the Ponzano-Regge amplitude for that triangulation as a function of the spins $(j_e)_{e\in\Gamma}$, we would like to derive a recursion relation under shifts of the spins $(j_e)_{e\in\Gamma}$. To approach the problem, it has been shown in \cite{pr3} that we can write this amplitude as a group integral.

We start by choosing a framing for the graph $\Gamma$, i.e. we embed it faithfully in a closed surface in $\Delta$. More precisely we choose a set of triangles forming a closed surface $\Sigma$, such that the edges of $\Gamma$ are all in $\Sigma$. In general, $\Gamma$ is not a triangulation of the whole surface $\Sigma$, but still defines a cellular decomposition of $\Sigma$ whose faces are delimited by the edges of $\Gamma$. We require that the faces are all homeomorphic to the 2-disk. Then the result proved in \cite{pr3} is that for a trivial topology\footnote{Reference \cite{pr3} also deals with the more complicated case of a non-trivial surface topology. With a non-trivial topology defined by the genus $g$ of the surface $\Sigma$, we need to associate group elements $U_C\in\SU(2)$ to each cycle $C$ of the surface. Assuming that each edge $e\in\Gamma$ belongs to a single cycle $C(e)$ (or to none), we have
\be
Z_\Delta[j_{e\in\Gamma}]=\int [\prod_fdg_f]\,\int [\prod_C dU_C]\,
\prod_e \chi_{j_e}(g_{s(e)}U_{C(e)}^{\epsilon_C(e)}g_{t(e)}^{-1})\;,
\ee
where the sign $\epsilon_C(e)$ records the relative orientation of the edge $e$ with respect to the orientation of the cycle.} $\Sigma\sim\cS^2$, the amplitude has a simple expression, with group elements $(g_f)$ attached to the faces of $\Sigma$,
\be
Z_\Delta[j_{e\in\Gamma}]=\int [\prod_fdg_f]\,
\prod_e \chi_{j_e}(g_{s(e)}g_{t(e)}^{-1})\;.
\ee
Here $s(e)$ and $t(e)$ denote the two faces of $\Sigma$ sharing the edge $e$. Since characters are invariant under taking the inverse of the argument, it does not matter which face we choose as the source or the target for each edge.

Written as such, this amplitude is the generalization of the integral for the squared 6j-symbol \eqref{int6jsquare}. If we look at the graph dual to $\Gamma$ in $\Sigma$, we see that the group elements $(g_f)$ are actually attached to the vertices of the dual graph. Thus it is the square of the evaluation of the spin network state dual to the graph $\Gamma$ on the surface $\Sigma$.

The next step is to re-express the measure using the class angles $\theta_e$ of $g_{s(e)}g_{t(e)}^{-1}$ as basic variables,
\be
Z_\Delta[j_{e\in\Gamma}]=\int [d\theta_{e}]^{\vert \Gamma\vert}\ \mu(\theta_{e})\,\prod_{e\in\Gamma} \sin\left(2j_e+1\right)\,\theta_e\;,
\ee
where $\mu$ is the Jacobian of the change of variables. Its analysis has been started in \cite{Feyndiag} where it has been proven that $\mu$ involves a product of Gram determinants similarly to the case of the squared 6j-symbol. However, the squared 6j-symbol is a special case where all the angular variables $\theta_e$ are independent. Generically that is not true because the elements $g_f$ are unit vectors in $\R^4$, so that only four of them can be linearly independent. Typically, the integral \eqref{10jgroup} for the 10j-symbol contains five copies of $\SU(2)$ which implies that one among the ten angles is determined by the others. That is the reason which leads $\mu$ to be a constraint in \eqref{10jangles}.

In the absence of constraints, one can use the method developed in the present paper to get recursion formulae. In the presence of a constrained measure $\mu$, it is in principle possible to also apply the method which was used on the 10j-symbol, provided that the constraints in $\mu$ are polynomials in $(\cos\theta_e)$. However, the geometric content of $\mu$ seems to depend on the connectivity matrix of $\Gamma$, so that at the end of the day, such recursion formulae lack a global, unified geometric interpretation.

%%%%%%%%%%%%%%%%%
\section{Outlook}
%%%%%%%%%%%%%%%%%

In this short paper, we have adapted a method outlined in \cite{recursion_simone} to derive recursion relations  for the square of the 6j-symbol using its expression as a multiple integral over $\SU(2)$. This method gives a very easy way to obtain recursion relations, which has a simple algebraic origin in the expression of the integral measure. Moreover, we have shown how it is simply related at leading order in the semi-classical limit to the closure of the tetrahedron whose quantum edge lengths are given by the spins labeling the 6j-symbol. Finally, we have explained how our method can be generalized to derive recursion relation for arbitrary spin foam amplitudes in the Ponzano-Regge model. We believe that this opens the door to a few issues and possibilities, which we hope to study in future work:
\begin{itemize}
\item Since the leading order of the recursion relation has a simple geometrical interpretation as the closure of the corresponding classical tetrahedron, we can hope that the next-to-leading order of our new recursion relation have a similar straightforward interpretation. It may also be that its analysis is technically less involved that when using the standard recursion as done in \cite{recursion_maite}.
\item It would be interesting to pursue this approach and further study the measure of the group integral for arbitrary spin foam amplitudes, in order to understand what geometric properties can emerge.
\item The existence of recursion relations in spin foam amplitudes is symptomatic of the existence of a symmetry. The standard recursion relation is usually related to the topological invariance of the Ponzano-Regge model under Pachner moves. Since our recursion relation can be interpreted as deformations of the space-time triangulation, it would be interesting to investigate its relation to discrete diffeomorphisms.
\item The specific invariance of 3d gravity and more generally BF theory in higher dimensions has been recently revisited \cite{PR_john, twisted-homology, matteo3}. In these works, one considers deformation properties of the full amplitude, using the notion of flat discrete connections, but not in the spin representation. That would be highly valuable to understand how the different objects used in these two different approaches actually probe the same geometric features.
%\item Our method applies in the present paper to 3d spin foam model and it would be interesting to see if it leads to a general method to generate similar recursion relations for arbitrary 4d spin foam models.
\end{itemize}

%%%
\section*{Acknowledgments}
%%%

EL is partially supported by the ANR ``Programme Blanc" grants LQG-09.\\
Research at Perimeter Institute is supported by the Government of Canada through Industry Canada and by the Province of Ontario through the Ministry of Research and Innovation.

%%%%%%%%%%%%%%%%%%%%%%%%%%%%%%%%%%%%%%%%
%\appendix

%%%%%%%%%%%%%%%%%%%%%%%%%%%%%%%%%%%%%%%%%%%%%%%%%%%%%%%%%%%%%%%%%%%%%%%%%%%%
%%%%%%%%%%%%%%%%%%%%%%%%%%%%%%%%%%%%%%%%%%%%%%%%%%%%%%%%%%%%%%%%%%%%%%%%%%%

\end{document}